\documentclass[aip,reprint]{revtex4-1}

\usepackage{graphicx}
\usepackage{amsmath}

\newcommand{\sI}{{\scriptscriptstyle{\mathrm I}}}

\begin{document}

\title{Locating the source of projectile fluid droplets}

\author{Christopher R.\ Varney}
\author{Fred Gittes}
%\email[]{gittes@wsu.edu}

\affiliation{Department of Physics and Astronomy, Washington State University,
Pullman, Washington 99164-2814}

\date{\today}

\begin{abstract}
The ill-posed projectile problem of finding the source height from spattered
droplets of viscous fluid is a longstanding obstacle to accident reconstruction
and crime scene analysis. It is widely known how to infer the impact angle
of droplets on a surface from the elongation of their impact profiles. However,
the lack of velocity information makes finding the height of the origin from
the impact position and angle of individual drops not possible.  From aggregate
statistics of the spatter and basic equations of projectile motion, we introduce
a reciprocal correlation plot that is effective when the polar launch angle
is concentrated in a narrow range.  The vertical coordinate depends on the
orientation of the spattered surface, and equals the tangent of the impact
angle for a level surface.  When the horizontal plot coordinate is twice the
reciprocal of the impact distance, we can infer the source height as the slope
of the data points in the reciprocal correlation plot.  If the distribution
of launch angles is not narrow, failure of the method is evident in the lack
of linear correlation. We perform a number of experimental trials, as well
as numerical calculations and show that the height estimate is insensitive
to aerodynamic drag. Besides its possible relevance for crime investigation,
reciprocal-plot analysis of spatter may find application to volcanism and
other topics and is most immediately applicable for undergraduate science and
engineering students in the context of crime-scene analysis.
\end{abstract}

%\pacs{9999}


\maketitle

\section{Introduction}
\label{Introduction}

The impact of spattered droplets of viscous fluid\cite{Vogel,Leyton} on a
horizontal surface results in elongated impact profiles, from which it is
easy to locate the vertical axis of origin from the orientation of elongated
impact profiles (see Fig.~1). It is accurate for blood-like
fluids (and routine practice in forensics\cite{Carter,Knock}) to estimate the
angle of impact of a viscous droplet from its impact profile as
\begin{equation}
\theta_\sI = \sin^{-1}\! \Big(\frac{\mathrm{profile\;width}}{\mathrm{length}}\Big).
\label{angle estimate}
\end{equation}
The subscript ``I'' denotes impact-related quantities. This construction,
which assigns to each elliptical profile (see Fig.~1, inset) the
projected shape of the incoming spherical droplet on the surface, works
remarkably well. From estimates of $\theta_\sI$ from Eq.~(\ref{angle
estimate}) we would like to estimate the location of the droplet launch
from the positions and angles of the impacts.

However, even when its vertical axis is known, the height of a source is not
deducible from the location and impact angle of individual drops ($r_\sI$ and
$\theta_\sI$ in Fig.~1). The impact velocity of each drop is missing the
information needed to infer the source height, and thus an height must
involve assumptions. A method widely used in forensics\cite{Carter} is to
extrapolate straight lines from each impact back to the launch axis and seek
a minimum in these height values, which is consistent only with the equations
of projectile motion in the limit of high velocity and low
aerodynamic drag. Otherwise, the strategy systematically overestimates the
source height\cite{Carter} (see Fig.~2). Such considerations have led to
alternative proposals, such as seeking missing velocity information in the
detailed structure of impact profiles.\cite{Knock}

In this paper we present a statistical and graphical method of back-estimation
consistent with the equations of projectile motion and provide
a height estimate in cases where the droplets are launched
within a narrow range of the launch angle, using only the impact
location and inferred impact angle. In the simplest case,
linearity appears in a plot of the tangent of the impact angle versus
twice the reciprocal distance of impact.

\begin{figure}
\includegraphics[width=1.0\columnwidth]
{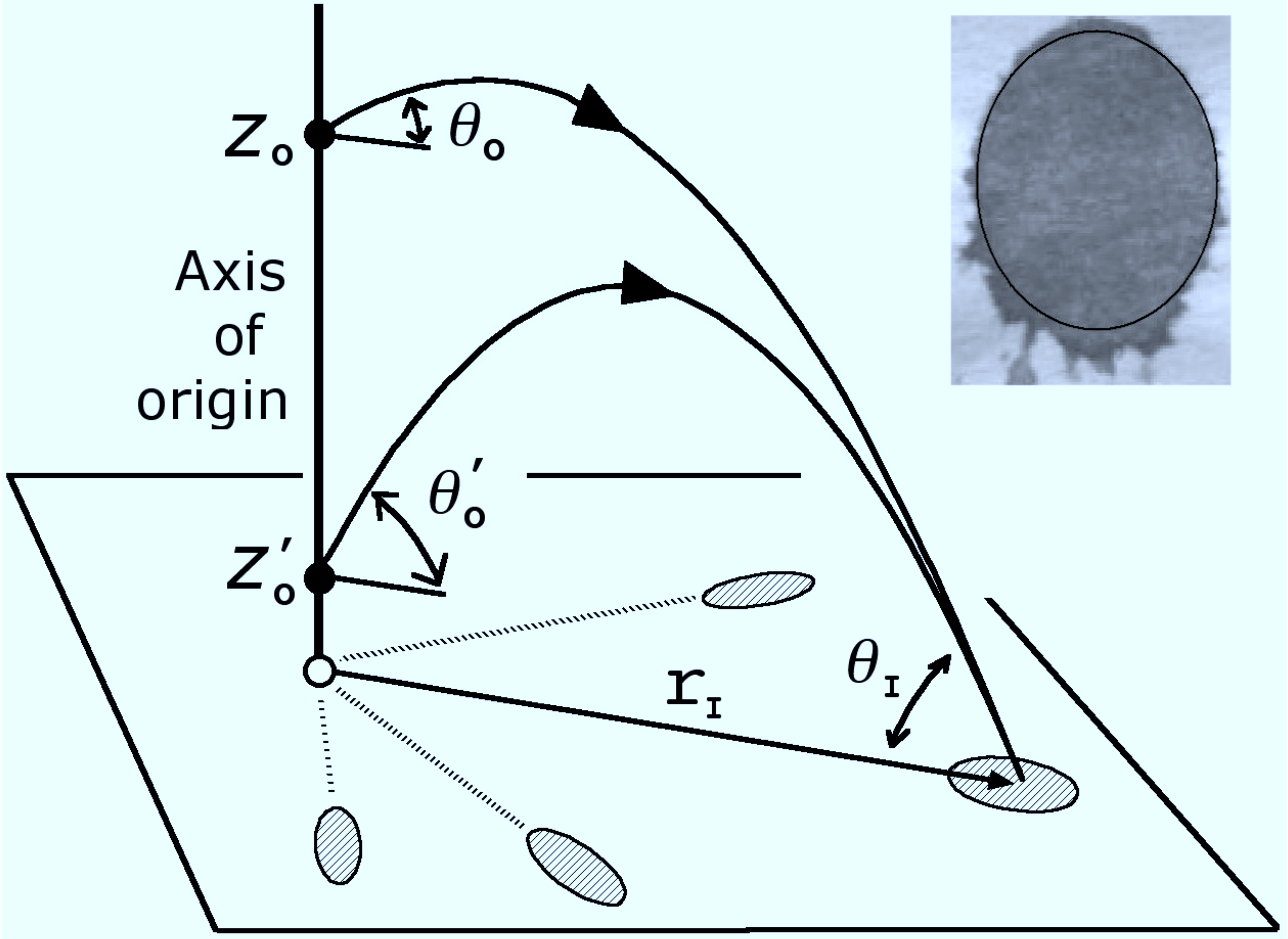}
\caption{
Profile orientations can be extrapolated (dotted lines) to a horizontal point
of convergence ($\circ$) so that the vertical axis of origin and the distance
$r_\sI$ are known. Even with knowledge of the impact angle $\theta_\sI$, we
still cannot determine the source height (for example, $z_0$ or $z'_0$). (Inset) Digitized
photograph of experimental impact profile fit to an ellipse. The impact angle is obtainable from the residue profile width and
length by attributing the elliptical profile to the projected shape of the
incoming spherical droplet.}
\label{Figure01}
\end{figure}

\begin{figure}
\includegraphics[width=1.0\columnwidth]
{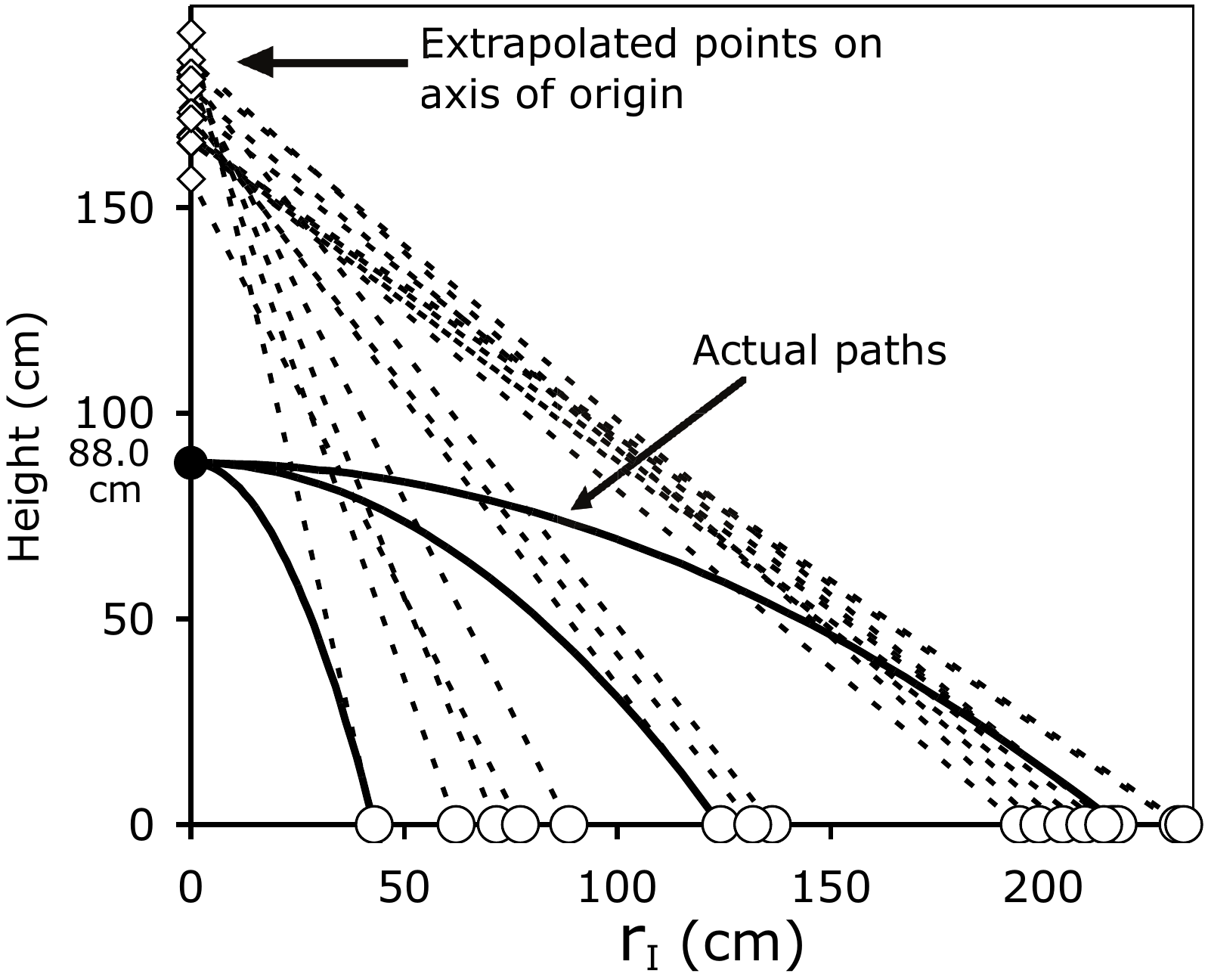}
\caption{
Position and impact-angle data ($\circ$) for floor spatter generated as
described in Section \ref{Methods} (clapper horizontal; i.e.\ launch angle
$0^\circ$).  Extrapolated points on the axis of convergence ($\diamondsuit$)
fail to estimate the source position ($\bullet$) (height $88.0\,\mathrm{cm}$)
even as a lower bound (a high-velocity assumption).
}
\label{Figure02}
\end{figure}

\begin{figure}
\includegraphics[width=0.76\columnwidth]
{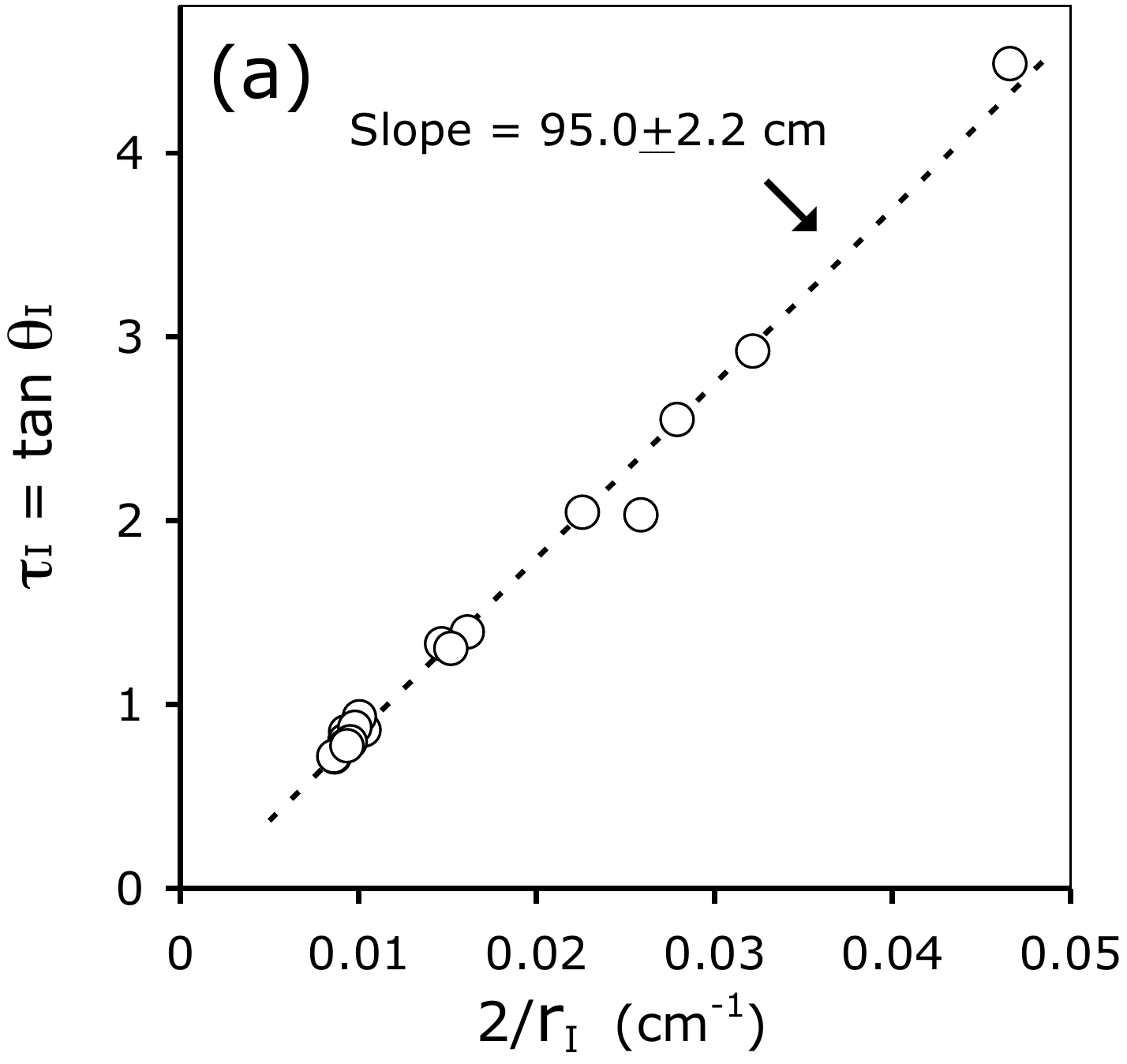}

\vspace{0.2cm}

\includegraphics[width=0.7\columnwidth]
{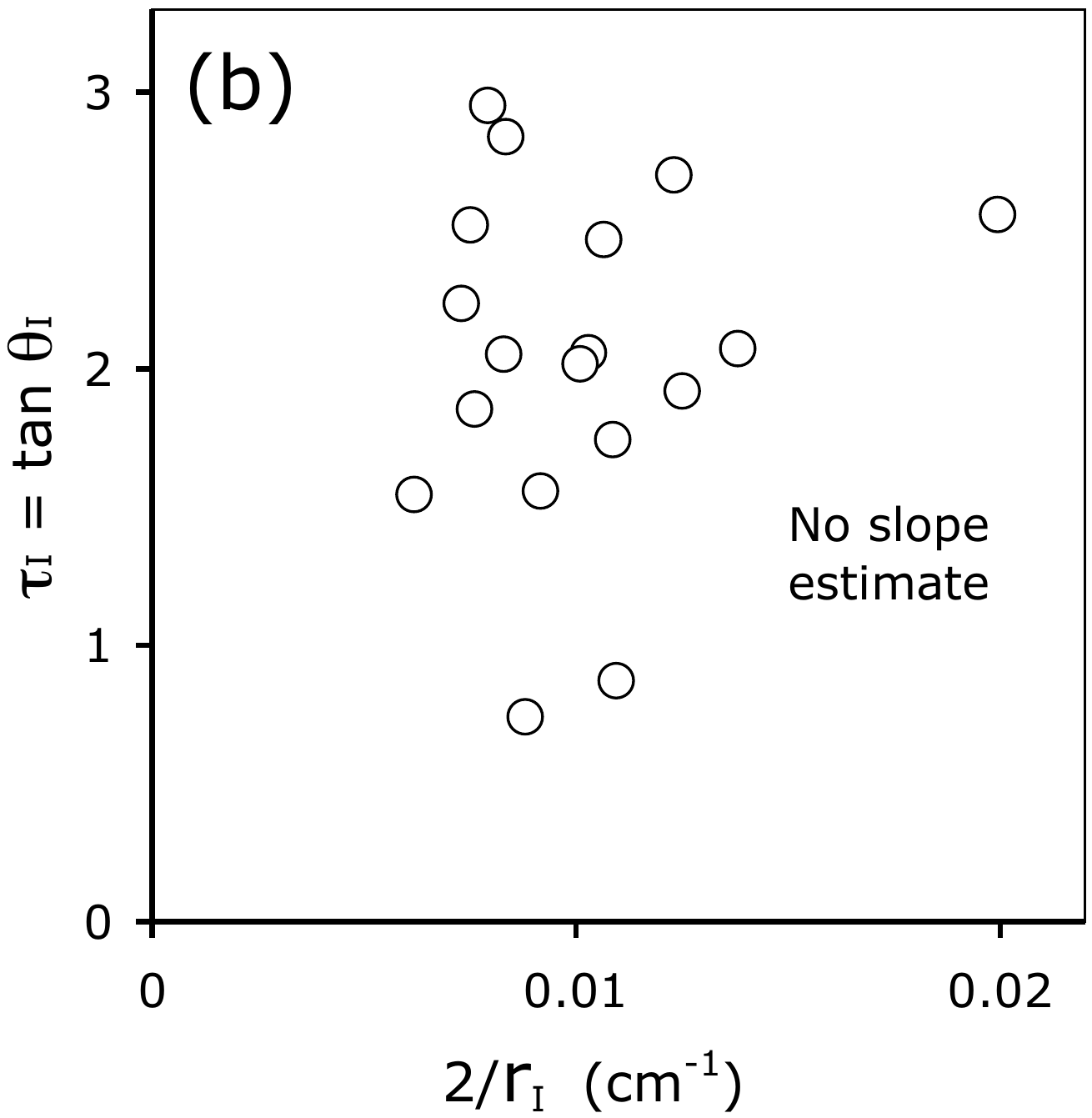}
\caption{(a)
The floor spatter data of Fig.~2 in a reciprocal correlation
plot, using $\tau_\sI= \tan\theta_\sI $ from Eq.~(\ref{tau-floor}). The slope
yields an estimated launch height of $95.0 \pm 2.2 \,\mathrm{cm}$ and an
estimated launch angle of $-5.8 \pm 2.4^\circ$. Pearson's coefficient of
linear correlation is $r=0.996$. The actual source height
is $88.0\,\mathrm{cm}$.
(b)
Trial with launch height $85.0\,\mathrm{cm}$ with the clapper sideways for a maximally broad launch angle distribution. In this case the
reciprocal plot yields no height estimate, as evidenced by lack of linear
correlation ($r=0.143$).}
\label{Figure03}
\end{figure}

\section{Reciprocal plot for floor spatter}
\label{Reciprocal plot}

The kinematic constant-acceleration
equations\cite{Serway,Knight} for a projectile droplet in cylindrical coordinates $z$ and $r$ (neglecting aerodynamic drag) are
\begin{equation}
z(t) = z_0 + \tfrac{1}{2} (v_z + v_{z,0}) t,
\label{const-a eqn}
\end{equation}
where $z_0$ is the actual launch height. Along the trajectory the horizontal
position is $r(t) = v_r t$ with $v_r$ the radial velocity. The vertical
velocity is $v_z = v_r\tan\theta$ with $\theta$ upward from the horizontal.
If we substitute $v_z = v_r\tan\theta$ and $v_{z,0} = v_r\tan\theta_0$ into Eq.~(\ref{const-a eqn}), we find
\begin{equation}
\tan\theta_\sI = (z_0 - z_\sI) \, \frac{2}{r_\sI} + \tan\theta_0.
\label{const-a recast}
\end{equation}
We have defined $\theta_\sI$ downward (that is, as $-\theta$) so that $\theta_\sI$
is positive at impact as in Fig.~1. For launch and impact at
equal height ($z_0=z_\sI$) Eq.~(\ref{const-a recast}) correctly predicts that
$\theta_0= \theta_\sI$. Consider impacts on a surface at $z_\sI=0$ (a floor),
where we have located the vertical axis of the origin (see Fig.~1)
and wish to find the launch height $z_0$. In Fig.~3(a) we plot the quantity for
actual spatter data,
\begin{equation}
\tau_\sI({\mathrm{floor}}) = \tan\theta_\sI
\label{tau-floor}
\end{equation}
versus the quantity $2/r_\sI$.
According to Eq.~(\ref{const-a recast}) our data will satisfy
\begin{equation}
\tau_\sI = z_0 \, \frac{2}{r_\sI} + \tan\theta_0.
\label{horiz RG}
\end{equation}
The slope, that is, the coefficient of $2/r_\sI$, gives the launch height $z_0$.
Figure~3(a) shows a trial using a standard ``clapper'' mechanism
(see Sec.~\ref{Methods}) in which a reciprocal correlation plot yields the
source height as $95.0 \pm 2.2$\,cm compared to the actual height of
88.0\,cm. The remaining discrepancy may indicate residual systematic
error in our procedure, such as a surface effect on profile shape or some
other factor.

\begin{figure}
\includegraphics[width=0.74\columnwidth]
{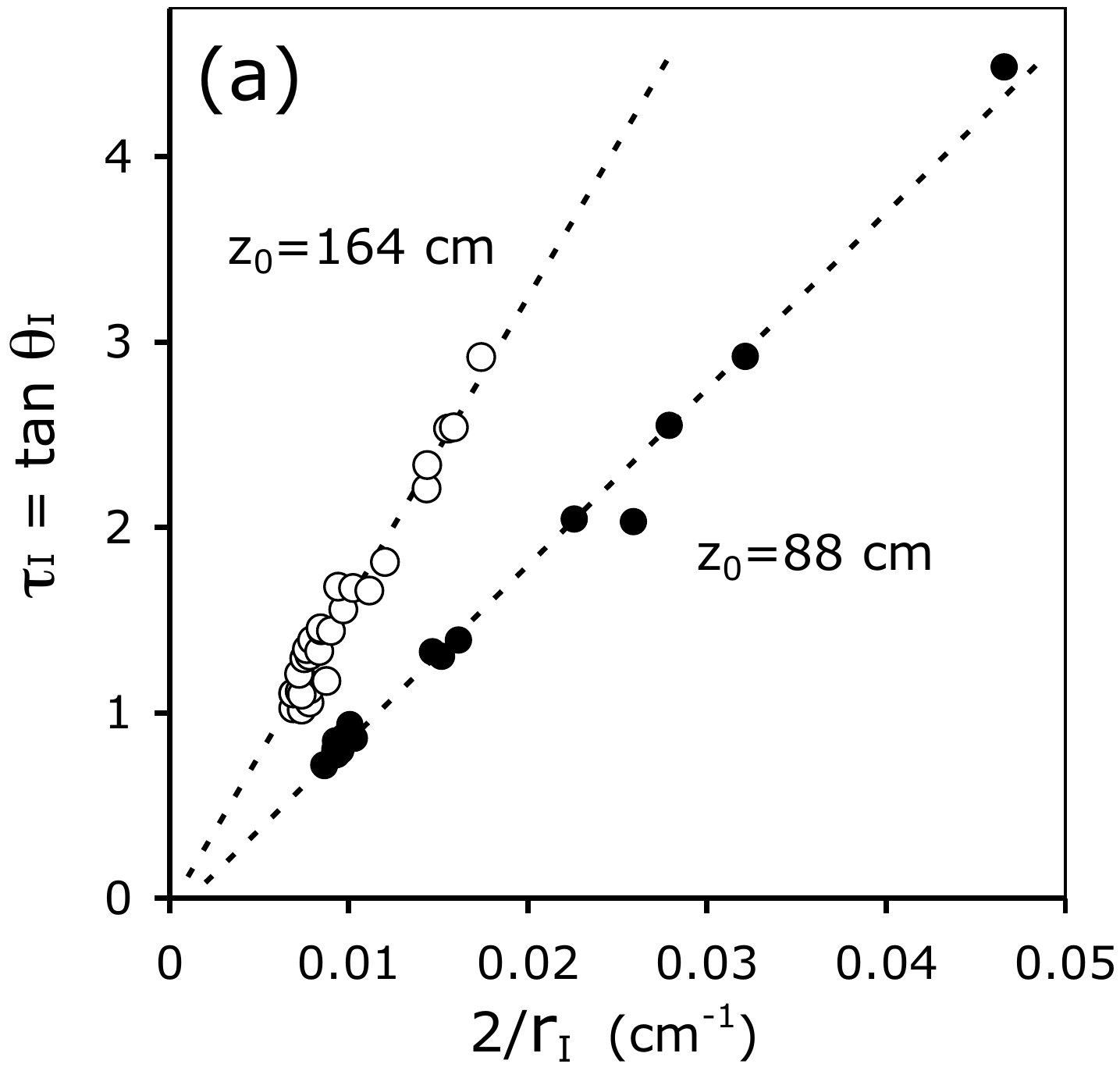}
\includegraphics[width=0.73\columnwidth]
{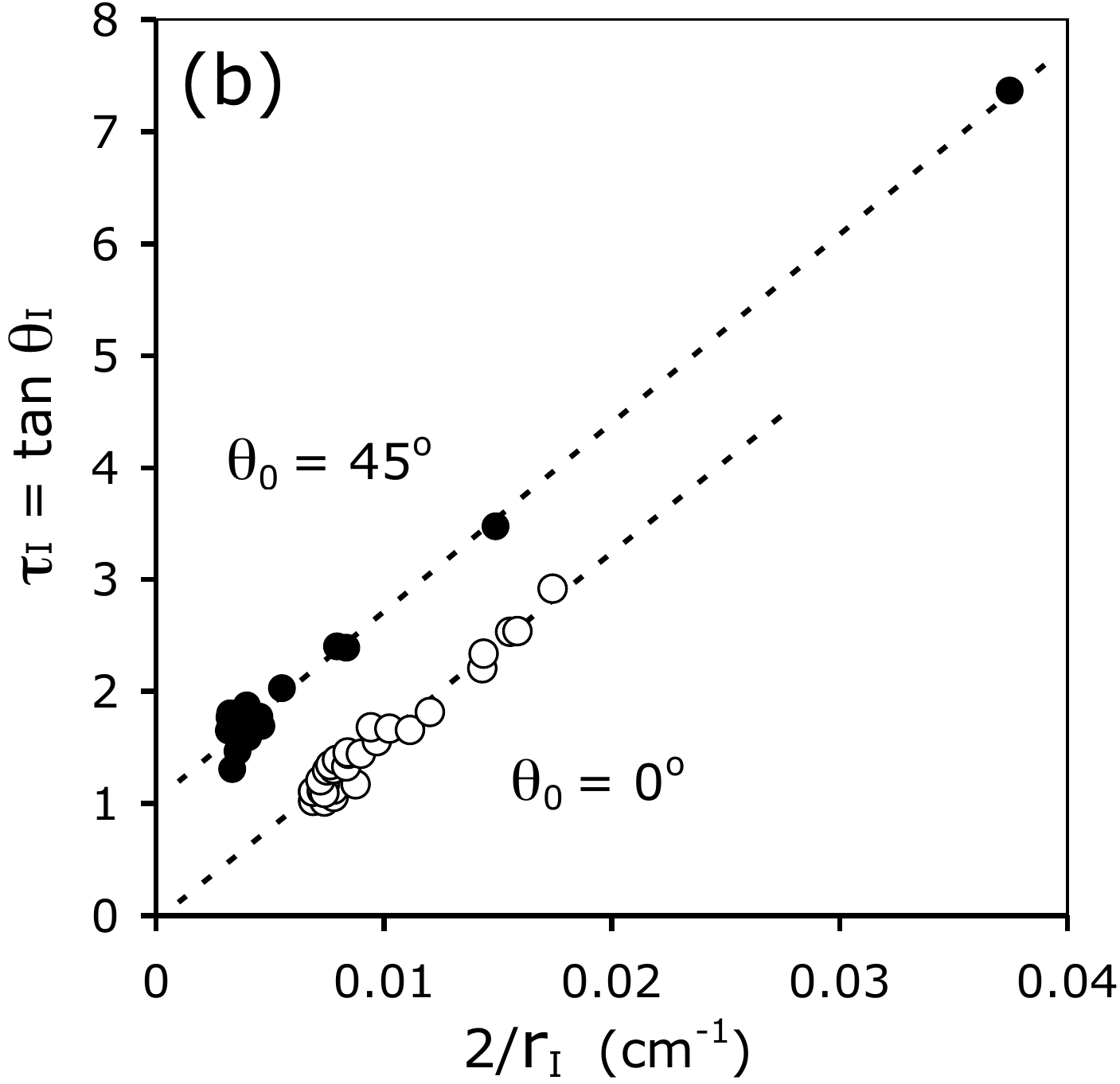}
\caption{
(a) Reciprocal correlation plots for floor impacts with horizontal launch:
($\bullet$), launch height $z_0= 88$\,cm (the same data as in
Fig.~2(b)); ($\circ$),
launch height $z_0= 164$\,cm ($\circ$). 
The clapper orientation is horizontal for both trials.
Note the differing slopes, due to their differing heights, but similar intercepts.
(Upper curve: slope $164\pm 6$\,cm, intercept $\theta_0 = 2.7\pm 3.7^\circ$, $r=0.996$. Lower curve as in Fig.~3(a).)
(b) Plots for floor impacts, with the clapper at $\theta_0=45^\circ$
($\bullet$) and $\theta_0=0^\circ$ ($\circ$), same data as in (a)). The heights of
165\,cm and 164\,cm, respectively, are similar. Note the similar slopes, but
distinct intercepts. (Upper curve: slope $168\pm 4$\,cm, intercept $\theta_0
= 45.8\pm 1.2^\circ$, and $r=0.995$. Lower
curve as in part (a).)}
\label{Figure04}
\end{figure}

To further illustrate the predictions of Eq.~(\ref{horiz RG}), Fig.~4(a) shows how a varying source height above a level surface leads
to different slopes in a reciprocal correlation plot. Figure~4(b)
shows how varying the launch angle of spatter leads to different vertical
intercepts.

Fig.~5 applies a reciprocal plot to spatter data collected from walls and
ceiling, as well as from the floor. Extensions of
Eq.~(\ref{tau-floor}) necessary for this are discussed in 
Sec.~\ref{Methods}.

\begin{figure}
\includegraphics[width=0.75\columnwidth]
{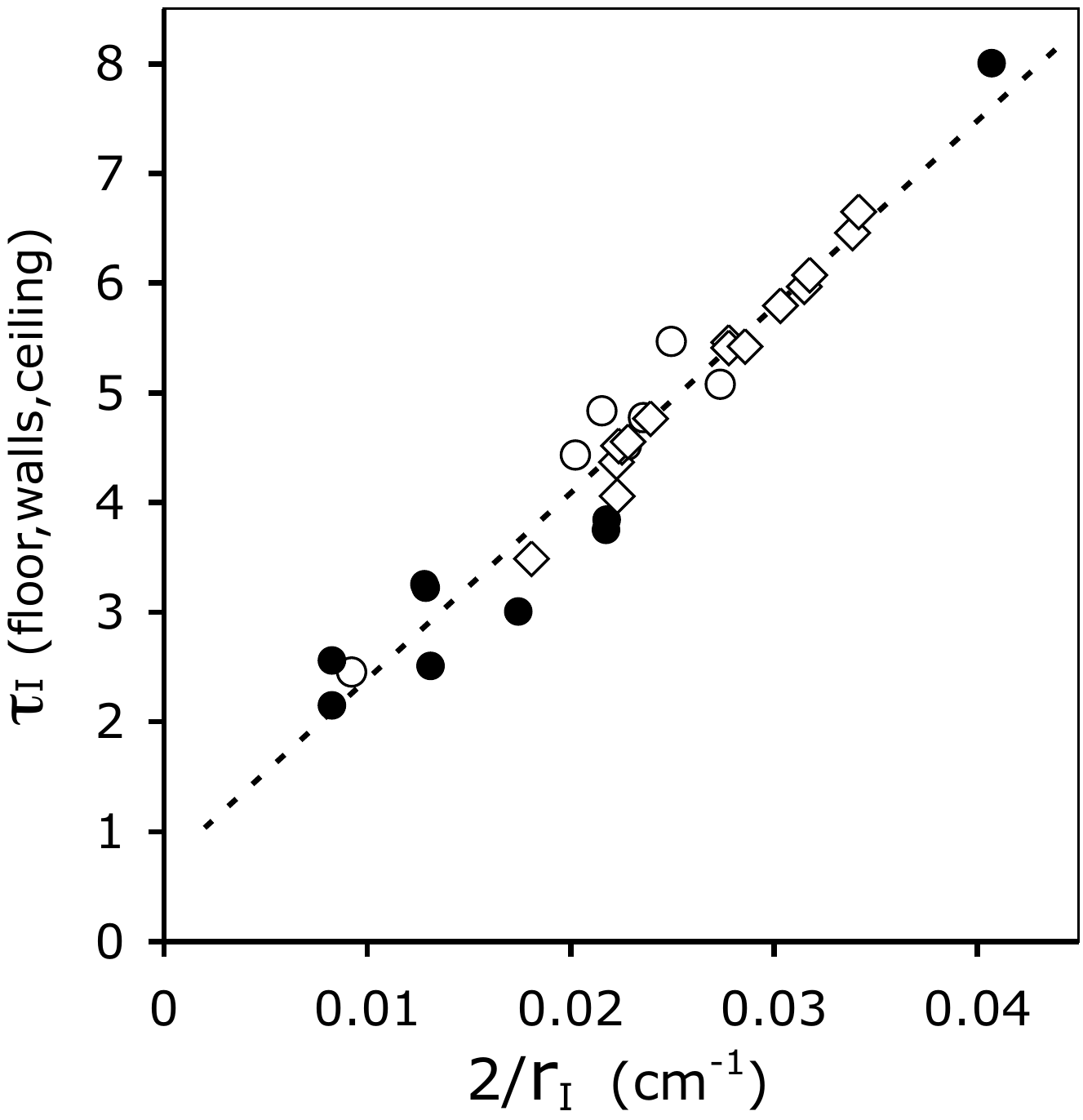}
\caption{
Reciprocal correlation plot of the combined data from the floor ($\circ$),
walls ($\bullet$), and ceiling ($\diamondsuit$) with $\tau_\sI$
generalized as described in Sec.~\ref{Methods}. For the combined
data the slope is $z_0 = 169 \pm 7 \,\mathrm{cm}$ and the intercept
is $35 \pm 7^\circ$ (dotted line), compared to the actual values of
$165\,\mathrm{cm}$ and $45^\circ$, with $r=0.973$. The separate slopes and intecepts of the floor
data (height $159 \pm 21 \,\mathrm{cm}$ and angle $48 \pm 13^\circ$),
wall data ($167 \pm 18 \,\mathrm{cm}$ and $34 \pm 14^\circ$), and
ceiling data ($185 \pm 14 \,\mathrm{cm}$ and $14 \pm 9^\circ$) are
not statistically distinguishable.}
\label{Figure05}
\end{figure}

\section{Aerodynamic drag}

In reciprocal plots such as Figs.~3(a) and 4,
points near the plot origin represent distant impacts, with high launch speed.
Here an error due to aerodynamic drag might be anticipated. Our trials
suggest no obvious problem, and therefore we turn to numerical calculations. The drag
force on a droplet with speed $v$ is\cite{Leyton,Batchelor,Vogel}
\begin{equation}
F_d = \frac{1}{2} \rho_\mathrm{air} A v^2 \,C_d,
\label{drag force}
\end{equation}
where $A$ is the cross-sectional area, $\rho_\mathrm{air}$ is the density of
air, and $C_d$ is an empirical coefficient. Separate calculations show that
for our spatter trials $C_d\approx 0.5$. We define
\begin{align}
v_{\scriptscriptstyle T} &= \sqrt{ \frac{4}{3} \frac{\rho_\mathrm{liq}}{\rho_\mathrm{air}} \frac{gd}{C_d}},\\
\noalign{\noindent and}
z_{\scriptscriptstyle T} &= \frac{v_{\scriptscriptstyle T}^2}{g},
\end{align}
where $g$ is the acceleration of gravity, $v_{\scriptscriptstyle T}$ is the
terminal velocity of a drop of diameter $d$ and density $\rho_\mathrm{liq}$,
and $z_{\scriptscriptstyle T}$ is a ``terminal height'' over which the terminal
velocity is nearly reached in a vertical fall ($v = 0.93\, v_{\scriptscriptstyle T}$ at $z_{\scriptscriptstyle
T}$). We write equations for the drop
trajectories which incorporate Eq.~(\ref{drag force}),
\begin{subequations}
\label{diffeqs}
\begin{align}
\frac{d \tilde v_x}{d \tilde t} & = -\tilde v \, \tilde v_x\\
\frac{d \tilde v_y}{d \tilde t} & = -(1 + \tilde v \, \tilde v_y)
\end{align}
\end{subequations}
where $\tilde v = (\tilde v_x^2 + \tilde v_y^2)^{1/2}$
and we use the dimensionless variables
\begin{equation}
\tilde v_{x,y} = \frac{v_{x,y}}{v_{\scriptscriptstyle T}},
\quad \tilde r_\sI = \frac{r_\sI}{z_{\scriptscriptstyle T}},
\quad \tilde t =\frac{v_{\scriptscriptstyle T}\,t}{g}.
\end{equation}
We numerically solved\cite{Press} Eq.~(\ref{diffeqs}) and constructed reciprocal plots in the presence of drag, at three launch angles
(see Fig.~4). These dimensionless plots, rescaled to the appropriate
$v_{\scriptscriptstyle T}$ and $z_{\scriptscriptstyle T}$, apply to any droplet
diameter or launch height. In all cases distortion is greatest on the left
side of the plot, as expected, where the range and launch speed are highest.

A notable feature of Fig.~6 is that, for moderate height and
launch speed where the plot is linear, aerodynamic drag mainly affects
intercepts, and thus the inferred launch angles, but not the slopes.
The height of origin estimates are only weakly affected. This dependence indicates a
certain robustness of the method in the presence of drag, as long as we
obtain linear plots, which might help explain the success of our plots in
Figs.~3(a), 4, and 5.

\begin{figure}
\includegraphics[width=0.9\columnwidth]
{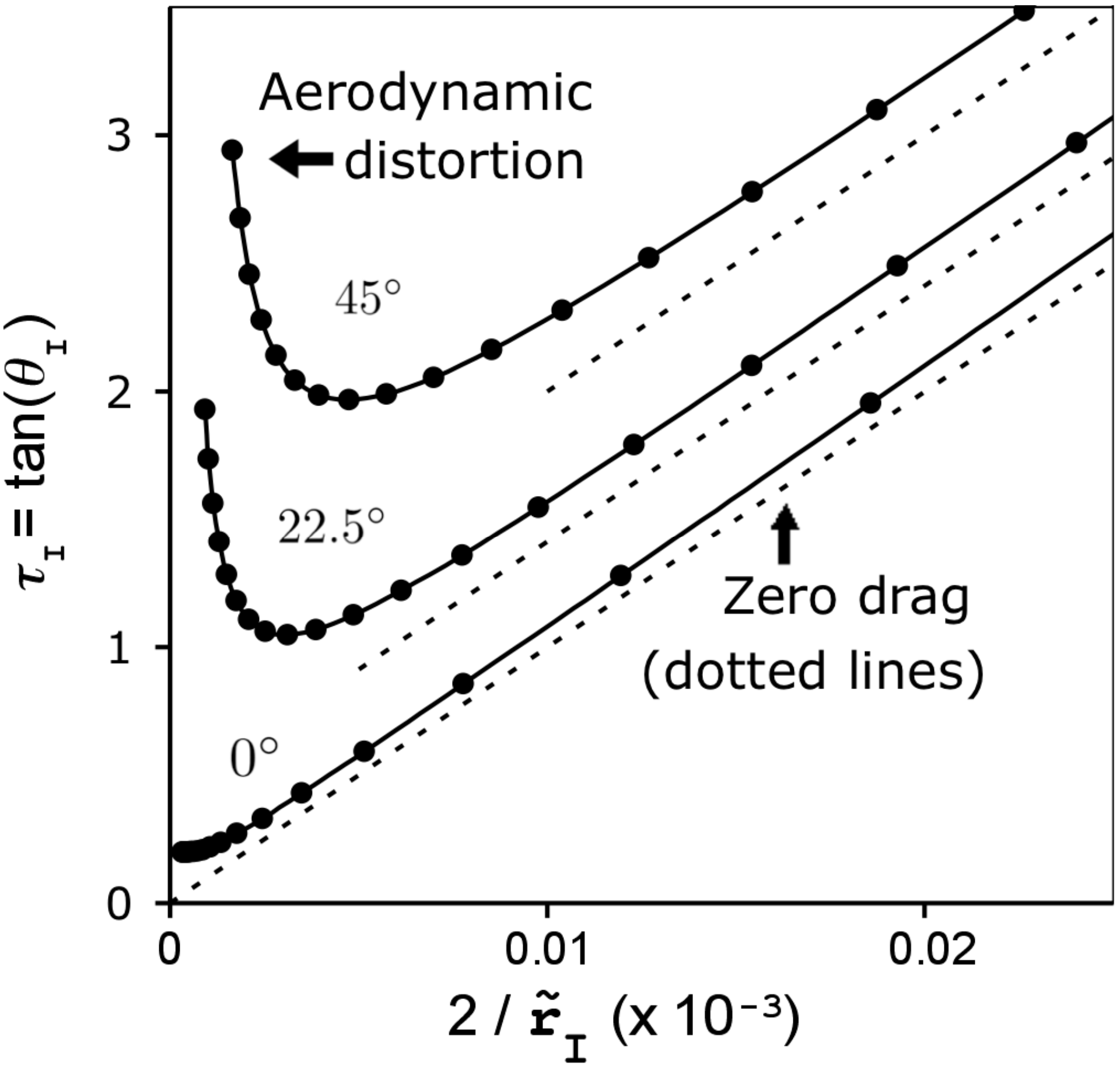}
\caption{
Reciprocal plots for trajectories in the presence of aerodynamic drag
($\bullet$) calculated using Eq.~(\ref{diffeqs}) at launch
angles $0^\circ$, $22.5^\circ$, and $45^\circ$. The source height is
$z_0 = 0.1 z_{\scriptscriptstyle T}$ and $\tilde r_\sI =
r_\sI/z_{\scriptscriptstyle T}$, where $z_{\scriptscriptstyle T}$ is
the characteristic fall height leading to the terminal velocity. The
horizontal axis is rescaled by $10^{-3}$ for rough correspondence to
cm$^{-1}$ in our experiments. In the linear region the slopes are little
changed from the straight lines (dotted) which would occur in the
absence of drag.}
\label{Figure06}
\end{figure}

\section{Methods and statistics}
\label{Methods}

In the spatter trials whose results are plotted in
Figs.~3--5, a viscous fluid was spattered on floors,
ceilings, and walls with a clapper device modeled after those used in forensics
training.\cite{Carter} Two wooden boards, joined at the rear by a spring-loaded
door hinge, slammed shut at the front, an impact area that we fitted with
metal plates. A small pouch of fluid to be spattered was taped to one 
plate.  The results in Figs.~3(a) and 4(a) were
obtained with the impact plates horizontal. To vary the launch angle we tilted the clapper forward and backward. For a
broad distribution of launch angles we rotated the
clapper about the long axis so that the launch area was vertically oriented.

We found that reliable impact angle estimates required considerable practice,
including self-calibration using vertical drop impacts on inclined surfaces,
during which we found that actual blood drops resulted in well-defined ellipses and
surprisingly good impact angle estimates using Eq.~(\ref{angle estimate}). To
avoid using actual blood in our spatter trials, we used a heterogeneous
fluid whose droplet impact profiles closely resemble those of blood droplets.
The majority of our experiments used approximately 2--3 parts Ashanti chicken
wing sauce (Bridge Foods) with 1 part 
Ivory dish soap to aid cleanup, and trace amounts of food coloring to enrich the color
for digitization. In other experiments, a viscous test fluid of 4 parts corn
syrup to 1 part water with food coloring also led to a reasonably successful
analysis.

From a typical spatter trial we chose approximately twenty profiles having well-defined elliptical profiles, representing a range of distances. For
the plots, we took digital photographs of the profiles, and used image analysis
software\cite{ImageJ} to manually fit the ellipses to the profile outlines shown in Fig.~1 (inset). We obtained best results by
matching well-defined portions of the elliptical outline. Very close to the
source, near-circular profiles can lead to large uncertainties both in Eq.~(\ref{angle estimate}) and in $2/r_\sI$ when $r_\sI$ is very small,\cite{Knock}
and hence some of these points were discarded.

Our estimates of slopes, intercepts, and their errors were made using standard
regression techniques,\cite{Press} where, for example, the estimate of the slope
(that is, launch height) is
\begin{equation}
z_\mathrm{est} = z_0 \frac{\langle{\Delta u\Delta\tau_\sI}\rangle }{\langle{(\Delta u\,)^2}\rangle}, \quad u = \frac{2z_0}{r_\sI},
\label{regression}
\end{equation}
with $z_0$ equal to the actual launch height and $\Delta u = u -\langle u \rangle$,
where $\langle\cdots\rangle$ denotes an average. A widely known
parameter of linear correlation\cite{Press} is Pearson's $r$, which for our
reciprocal plots is
\begin{equation}
r = \frac{\langle \Delta u\Delta \tau_\sI \rangle}{ \sqrt{\langle{(\Delta u )^2}\rangle\langle{(\Delta \tau_\sI)^2}\rangle }}.
\label{Pearson}
\end{equation}
A complete description of projectile motion requires that Eq.~(\ref{const-a
recast}) be combined with the relation
\begin{equation}
\tan^2\theta_\sI = (\eta^2\!+1)\tan^2\theta_0 + \eta^2, \label{quadratic}
\end{equation}
which parameterizes lines of constant $\theta_0$ in the reciprocal plot by
the variable $\eta = \sqrt{2gz_0}/v_0$, where $v_0$ is the launch
speed.

Consider the special case of a product probability distribution for the
launch angle and velocity $P(\theta_0,v) = P_1(\theta_0)P_2(v)$, where
$P_1(\theta_0)$ is symmetrical about the horizontal. We can obtain, using Eqs.~(\ref{const-a recast}), (\ref{regression}), (\ref{Pearson}), and (\ref{quadratic})
the simple relation
\begin{equation}
\label{thisresult}
z_\mathrm{est} = r^2 z_0\ \text{[for symmetric $P_1(\theta_0)$]}.
\end{equation}
Although Eq.~\eqref{thisresult} is not general, it suggests that bias may occur in height
estimates if $r^2$ is not reasonably close to $1$. A requirement of strong
linear correlation also accords with common sense. Figure~3(b)
showed an unfavorable trial in which the spread in launch angle is maximized
by turning the clapper sideways. This trial was performed ten times,
yielding the $r$ values $0.14$, $0.18$, $0.78$, $0.86$, $0.80$, $0.56$,
$0.83$, $0.58$, $0.72$, and $0.37$.
These values are all smaller
than $1$, and the plots were uncorrelated in appearance. In contrast, in Figs.~3(a)
and 4 $r=0.995$ or higher, and for
the combined data sets in Fig.~5, $r=0.973$.

We can also combine and analyze spatter data from floors and other
non-horizontal surfaces in a single reciprocal plot.
Co-plotting data from non-horizontal surfaces (see Fig.~5) requires
corresponding redefinitions of the variable $\tau_\sI$. We then seek a slope
via Eq.~(\ref{horiz RG}) as before. Spatter on a ceiling at height $z=z_\sI$
is incorporated by adding a term to Eq.~(\ref{tau-floor}) and changing the sign
of the tangent,
\begin{equation}
\tau_\sI({\mathrm{ceiling}})
= z_\sI \frac{2}{r_\sI} - \tan\theta_\sI.
\label{tau-ceiling}
\end{equation}
Ceiling data tended to be more robust closer to the source. On walls the
right-hand side of Eq.~(\ref{angle estimate}) yields the angle
$\theta_{\scriptscriptstyle W}$ of impact with the plane of the wall. The
correct vertical angle is then $\theta_\sI =\sin^{-1} \left(\cos\alpha_\sI
\cos\theta_{\scriptscriptstyle W}\right)$ with $\alpha_\sI$ the angle from
vertical of the wall profile. We plot
\begin{equation}
\tau_\sI({\mathrm{walls}}) = z_\sI \frac{2}{r_\sI} + \tan\theta_\sI,
\label{tau-wall}
\end{equation}
where $z_\sI$ is the elevation of the impact. We typically avoided near-horizontal
wall impacts. Once the axis of origin is located, $\tan\theta_\sI$ in Eq.~(\ref{tau-wall}) can be replaced by $\cos\phi_\sI/\tan\alpha_\sI$, where $\phi_\sI$
is the angle between the wall and the vertical plane of impact, a relation which
is convenient for curved vertical surfaces such as pipes.

\section{Discussion}
\label{Discussion}

We have introduced and illustrated with several experiments, a
plot-based method for locating the spatial source of spattered viscous fluid. Th method is effective when the spatter is launched within a narrow range of
polar angles. We showed how a reciprocal
plot of the impact data, together with elementary projectile physics, can exhibit
linear trends among the data points. From the slope of any strong linear
correlation that occurs, we obtain a robust estimate of the height of the
origin that would otherwise be unavailable. Broad distributions of launch
angle cause the method to fail, and in such cases we reach a null
conclusion (a lack of linear correlation), rather than an erroneous estimate
of height of origin.

We also showed that the method appears insensitive to aerodynamic drag effects within some velocity
regime. We also extended the plotting method to some other simple geometries.

The reciprocal plot, which is based on the correct equations of projectile motion,
may eventually become a useful tool for forensic spatter analysis, where
information on height of spatter origin pertains, for example, to whether a
victim was sitting or standing.

Another possible use of this analysis, which we are pursuing separately, may
be its application to geophysics and volcanism, especially to phenomena such
as volcanic ejection and lava fountains.\cite{Wolff}

The most immediate use of reciprocal-plot analysis is as an
undergraduate activity for science and engineering students. In the simplest
case students might use a wooden block or book to spatter a puddle of viscous
``blood'' (such as corn syrup and food coloring) from an elevated surface
such as a table or shelf, tabulate the position and eccentricities of droplet
profiles on paper taped to the floor, and analyze this data via reciprocal
plots in a spreadsheet program to reproduce our Fig.~3(a). Such
an activity is a novel use of the equations of
projectile motion within the exciting context of crime-scene investigation.

\begin{acknowledgments}
The authors acknowledge valuable initial discussions with Dr.\ Christopher
Dudley. We are grateful for the generous assistance of Dr.\ Gabriel Hanna.
We are particularly indebted to Professor Anita Vasavada for helping to bring
the work into its final form, and to reviewers of our manuscript for valuable
advice. The authors also wish to thank Lake Chelan Community Hospital, in
particular, Mrs.\ Melissa Hankins, and Mr.\ Lee Reynolds for generous donations,
Dr.\ Katherine Taylor for her interest and suggestions, Dr.\ Fred Carter for
his input, Mr.\ Tom Johnson for materials and space graciously provided, and
Mr.\ Greg Varney and Mrs.\ Joanne Varney for materials provided.
\end{acknowledgments}

\end{document}